\documentclass{article}
\usepackage{epsfig}
\usepackage{amsmath}
\newtheorem{theorem}{Theorem}[section]
\newtheorem{lemma}{Lemma}[section]

\newcommand{\blackslug}{\penalty 1000\hbox{   \vrule height
8pt width .4pt\hskip -.4pt    \vbox{\hrule width 8pt height .4pt\vskip -.4pt           \vskip 8pt
\vskip -.4pt\hrule width 8pt height .4pt}    \hskip -3.9pt     \vrule height 8pt width .4pt}}

\newenvironment{proof}{\vspace{1mm} \noindent {\sc
Proof.}$\;$\rm}{\qed}
\newcommand{\qed}{\hspace*{\fill}\blackslug}
\def\boxit#1{\vbox{\hrule\hbox{\vrule\kern4pt
\vbox{\kern1pt#1\kern1pt}
\kern2pt\vrule}\hrule}}
\begin{document}

\title{\bf On the Independent Set and Common Subgraph Problems in Random Graphs}

\author{Yinglei Song \\
School of Computer Science and Engineering, Jiangsu University of Science and Technology\\
Zhenjiang, Jiangsu 212003, China\\
yingleisong@gmail.edu\\
}
\date{}
\maketitle

\begin{abstract}

\noindent In this paper, we develop
efficient exact and approximate algorithms for computing a maximum independent set in random graphs. In a
random graph $G$, each pair of vertices are joined by an edge with a probability $p$, where $p$ is a constant
between $0$ and $1$. We show that, a maximum independent set in a random graph that contains $n$ vertices can
be computed in expected computation time $2^{O(\log_{2}^{2}{n})}$. Using techniques based on enumeration,
we develop an algorithm that can find a largest common subgraph in two random graphs in $n$ and $m$ vertices ($m \leq n$) in
expected computation time $2^{O(n^{\frac{1}{2}}\log_{2}^{\frac{5}{3}}{n})}$. In addition, we show that, with
high probability, the parameterized independent set problem is fixed parameter tractable in random graphs and
the maximum independent set in a random graph in $n$ vertices can be approximated within a ratio of
$\frac{2n}{2^{\sqrt{\log_{2}{n}}}}$ in expected polynomial time.
\end{abstract}

\section{Introduction}
In computer science, many optimization problems can be reduced to the optimization of objectives that are
formulated and described in a graph. The development of efficient exact or approximate algorithms for graph
optimization problems thus constitute an important part of the research in combinatorial optimization.
However, a large number of graph optimization problems have been shown to be NP-hard \cite{garey}, which
suggests that it is unlikely to develop algorithms that can solve these problems in polynomial time. A well
known example is the {\sc Maximum Independent Set} problem. Given a graph $G=(V,E)$, a vertex set $I
\subseteq V$ is an {\it independent set} if there is no edge
between any pair of two vertices in $I$. The goal of the {\sc Maximum Independent Set} problem is to find an
independent set of the largest size in a given graph $G$. The problem can be trivially solved in time
$2^{O(n)}$ by enumerating and checking all possible vertex subsets in the graph. Although intensive research
has been performed to improve the computation time needed to find an optimal solution
\cite{balas,carraghan,fahle,fomin,tarjan,jian,konc,ostergard,pardalos,regin,robson,robson1,tomita,tomita1},
an algorithm that needs subexponential time is not yet available for this problem. Recently, it is proposed
that this problem is unlikely to be solved in subexponential time
\cite{chen,chen2}.

Due to the difficulty of developing efficient algorithms that can find optimal solutions for these problems, a
large number of algorithms have been developed to generate approximate solutions that are close to optimal
ones in polynomial time \cite{johnson}. Solutions provided by these algorithms are often guaranteed to be
within a ratio of the optimal solution and thus can be useful in practice. For example, the {\sc Minimum
Vertex Cover} problem can be approximated by a simple polynomial time algorithm within a ratio of $2.0$.
However, it has been shown that it is NP-hard to approximate this problem within a ratio of $1.362$
\cite{dinur}. A well known inapproximability result regarding the {\sc Maximum Independent Set} problem is
that it is NP-hard to approximate the maximum independent set in a graph within a ratio of $n^{1-\epsilon}$,
where $0 < \epsilon<1$ is a constant and $n$ is the number of vertices in the graph \cite{hastad}. This result
suggests that an approximate solution with a guaranteed constant approximate ratio cannot be obtained in
polynomial time for the {\sc Maximum Independent Set} problem unless NP=P. So far, the best known
approximation ratio that has been achieved for this problem in general graphs is
$O(\frac{n\log_{2}^{2}{\log_{2}{n}}}{\log_{2}^{3}{n}})$ \cite{fiege}.

For those problems that cannot be even approximated within a good approximation ratio in polynomial time, such
as the {\sc Maximum Independent Set} problem, heuristics that can efficiently generate approximate solutions
are often employed in practice to solve them \cite{battiti,katayama,grosso}. However, solutions
generated by heuristics are not guaranteed to be close to the optimal ones and their applications are thus
restricted to scenarios where the accuracy of solutions is not a crucial issue.

Parameterized computation provides another potentially practical solution for some problems that are
computationally intractable. In particular, one or a few parameters in some intractable problems can be
identified and parameterized computation studies whether efficient algorithms exist for these problems while
all parameters are small. A parameterized problem may contain a few parameters $k_1, k_2, \cdots, k_l$ and the
problem is {\it fixed parameter tractable} if it can be solved in time $O(f(k_1, k_2, \cdots, k_l)n^{c})$,
where $f$ is a function of $k_1, k_2, \cdots, k_l$, $n$ is the size of the problem and $c$ is a constant
independent of all parameters. For example, the {\sc Vertex Cover} problem is to determine whether a graph
$G=(V,E)$ contains a vertex cover of size at most $k$ or not. The problem is NP-complete. However, a simple
parameterized algorithm can solve the problem in time $O(2^{k}|V|)$ \cite{downey1}. In practice, this
algorithm can be used to efficiently solve the {\sc Vertex Cover} problem when the parameter $k$ is fixed and
small. On the other hand, some problems do not have known efficient parameterized solutions and are therefore
parameterized intractable. Similar to the conventional complexity theory, a hierarchy of complexity classes
has been constructed to describe the parameterized complexity of these problems \cite{downey1}. For example,
the {\sc Independent Set} problem is to decide whether a graph contains an independent set of size $k$ or not
and has been shown to be W[1]-complete \cite{downey2}. It cannot be solved with an efficient parameterized
algorithm unless all problems in W[1] are fixed parameter tractable. A thorough investigation on these
parameterized complexity classes are provided in \cite{downey}.

In this paper, we develop exact and approximate algorithms for the {\sc Maximum Independent Set} problem where
the underlying graph is a random graph generated based on the Erd\H{o}s R\'{e}nyi model
\cite{erdos}. Such a random graph is generated by treating each
pair of vertices independently and adding an edge to join them with a probability of $p$ ($0<p<1$), where $p$
is a constant. Recent research in molecular biology has shown that the protein side chain interaction network
conforms remarkably well to random graphs generated by the Erd\H{o}s R\'{e}nyi model \cite{brinda}. Therefore,
efficient algorithms for some NP-hard problems in random graphs, if exist, may significantly improve the
computational efficiency for some important optimization problems related to protein structure prediction.

In \cite{grimmett,karp}, it has been shown that with high probability, the maximum independent set in a random
graph is of size $O(\log_{2}{n})$. However, this result does not directly lead to an algorithm that can
compute the maximum independent set in a random graph in expected subexponential time. In \cite{fiege2}, a
polynomial time algorithm that can compute a maximum independent set in a sparse random graph with high
probability is developed. However, the algorithm is based on a large independent set that is embedded in the
graph and thus cannot be used for all graphs. We show that the maximum independent set in a random graph can
be computed in expected computation time $2^{O(\log_{2}^{2}{n})}$, where $n$ is the number of vertices in the
graph. This result significantly improves the best known time complexity $O(2^{\frac{n}{4}})$ for finding a
maximum independent set in general graphs \cite{robson1}.

Using techniques based on enumeration, we develop an algorithm that can compute a largest common subgraph of two random
graphs of $n$ and $m$ vertices ($n \geq m$) in expected computation time
$2^{O(n^{\frac{1}{2}}\log_{2}^{\frac{5}{3}}{n})}$. This result significantly improves on the best known time
complexity $2^{O(m\log_{2}{n})}$ for this problem when $m=O(n)$. In addition, we show that, with high
probability, the parameterized independent set problem is fixed parameter tractable in random graphs. For
approximate algorithms, we develop an algorithm that can achieve an approximation ratio of
$\frac{2n}{2^{\sqrt{\log_{2}{n}}}}$ in expected polynomial time, which is a significant improvement compared
with the best known approximate ratio that can be achieved in general graphs
\cite{abu,suters}.

\section{Maximum Independent Set in Random Graphs}
A {\it random graph} $G(V, p)$, where $0<p<1$, is a graph obtained by independently adding edges between each
pair of vertices in $V$ with a probability $p$. Given a vertex $v \in V$, the {\it degree} of $v$ in $G$ is
the number of vertices that are connected to $v$ by an edge in G. We use $deg_{G}(v)$ to denote the degree of
vertex $v$ in graph $G$ and $N_G(v)$ to denote the set of vertices that are connected to $v$ by an edge in
$G$. A vertex subset $I
\subseteq V$ is an independent set in $G$ if there is no edge
between any pair of vertices in $I$. The goal of the {\sc Maximum Independent Set} problem is to find an
independent set of the largest size in a given graph.

In \cite{grimmett,karp}, it is shown that, with high probability,
the size of a maximum independent set in a random graph $G(V,p)$ is
$\frac{2\log_{2}{n}}{\log_{2}{\frac{1}{1-p}}}$, where $n$ is the number of
vertices in $G$. A straightforward algorithm by exhaustively enumerating all
vertex subsets of size $\frac{2\log_{2}{n}}{\log_{2}{\frac{1}{1-p}}}$
can thus compute a maximum independent set in most random graphs in
time $n^{O(\log_{2}{n})}$. However, to compute a maximum
independent set in all random graphs, the
algorithm must be able to cope with the cases where the graph contains
an independent set of size larger than $O(\log_{2}{n})$. The
algorithm needs time $2^{O(n)}$ to compute a maximum independent set in these
cases. The best known upper bound
of the probability for a random graph to have
a maximum independent set larger than $O(\log_{2}{n})$ is $\frac{1}{n^{O(1)}}$ \cite{grimmett,karp},
the expected time complexity of this enumeration based algorithm is thus $2^{O(n)}$.

We show that the maximum independent set in a random
graph $G=(V, p)$ can be computed in expected subexponential time.

\begin{lemma}
\label{lm1}
\rm
Given a random graph $G=(V, p)$ where $n=|V|$ and a sufficiently small constant $\epsilon$ such that
$\epsilon<p$, there exists a vertex $v \in V$ such that $deg_{G}(v) \geq (p-\epsilon)n$ with probability at
least $1-2^{-\mu n^{2}}$, where $\mu$ is a positive constant that only depends on $\epsilon$ and $p$.

\begin{proof}
If such a vertex does not exist, the number of edges $n(E)$ in $G$ is at most $\frac{(p-\epsilon)n^{2}}{2}$
since the degree of each vertex is at most $(p-\epsilon)n$. However, from the construction of graph $G$, the
expected number of edges in $G$ can be obtained as follows
\begin{equation}
E(n(E))=\frac{pn(n-1)}{2}
\end{equation}
From Chernoff bound, we can bound the probability for $n(E)<\frac{(p-\epsilon)n^{2}}{2}$ by
\begin{equation}
Pr(n(E)<\frac{(p-\epsilon)n^2}{2})<\exp{(-\frac{pn(n-1)\delta^2}{4})}
\end{equation}
where $\delta=\frac{n\epsilon-p}{p(n-1)}$. For sufficiently large $n$, we have
\begin{eqnarray}
\delta & > & \frac{\epsilon}{2p} \\
n-1 & > & \frac{n}{2}.
\end{eqnarray}
We can thus immediately obtain
\begin{eqnarray}
Pr(n(E)& < & \frac{(p-\epsilon)n^2}{2}) \\
       & < & \exp{(-\frac{\epsilon^2n^2}{32p})} \\
       & = & 2^{-\frac{\epsilon^2n^2}{32p\ln{2}}}.
\end{eqnarray}
We then let $\mu=\frac{\epsilon^2}{32p\ln{2}}$ and we conclude that with probability at least $1-2^{-\mu
n^2}$, there exists vertex $v
\in V$ such that $deg_{G}(v) \geq (p-\epsilon)n$.
\end{proof}
\end{lemma}

The proof of Lemma \ref{lm1} relies on the fact that $p$ is a constant independent of $n$, the Lemma does not
hold if the value of $p$ depends on $n$. A random graph $G=(V,p)$ in $n$ vertices is {\it good} if it contains at least one vertex whose degree is at least $(p-\epsilon)n$. Given a random graph, the algorithm starts by finding a vertex $v$ such that
$deg_{G}(v)$ is at least $(p-\epsilon)n$. If such a vertex does not exist, the algorithm enumerates all
subsets of $V$ and returns an independent set of the largest size. If $v$ exists, the algorithm branches on
two possible cases on whether $v$ is contained in $I$ or not. In particular, if $v \in I$, $v$ and vertices in
$N(v)$ are deleted from $G$ and the resulting graph is $G_1$; if $v \notin I$, $v$ is deleted from $G$ and the
resulting graph is $G_2$. The algorithm is then recursively applied on both $G_1$ and $G_2$ to compute a
maximum independent set in each of them. We use $I_1$ and $I_2$ to denote the maximum independent sets in
$G_1$ and $G_2$ found by the algorithm respectively. $I_2$ is returned as a maximum independent set in $G$ if
$|I_2| \geq |I_1|+1$ and $I_1 \cup \{v\}$ is returned otherwise. We show that this algorithm terminates in
expected time $2^{O(\log_{2}^{2}{n})}$.

\begin{theorem}
\label{th1}
\rm
A maximum independent set in a random graph $G=(V,p)$ with $n$ vertices can be computed in expected
computation time $2^{O(\log_{2}^{2}{n})}$.

\begin{proof}
We show that the algorithm described above terminates in expected time $2^{O(\log_{2}^{2}{n})}$. In
particular, the algorithm is recursive and for each step of recursion, we have the following recursion
relation for the computation time if the underlying graph is good and contains $m$ vertices
\begin{equation}
T(m) \leq T((1-p+\epsilon)m)+T(m-1)+O(m^2)
\end{equation}
where $T(m)$ is the computation time needed by the algorithm in a graph on $m$ vertices. The term $O(m^2)$ is the
computation time needed to find a vertex whose degree is at least $(p-\epsilon)m$, since the time needed to
compute the degree of a vertex is $O(m)$ and the algorithm may need to check $m$ vertices to find such a
vertex. If the underlying graph is not good, the algorithm exhaustively enumerates all subsets in the graph
and finds an independent set of the largest size. The computation time is $2^{O(m)}$.

We are now ready to establish the expected computation time for the algorithm. In particular, we use $ET(m)$
to denote the expected computation time of the algorithm on a graph that contains $m$ vertices. From Lemma
\ref{lm1}, an underlying graph $G'$ in $m$ vertices is good with a probability of at least $1-2^{-\mu m^2}$.
We thus can immediately obtain the following recursion for $ET(m)$.
\begin{eqnarray}
ET(m) & \leq & ET((1-p+\epsilon)m)+ET(m-1)+O(m^2)+2^{O(m)-\mu m^2}
\\
      & \leq &  ET((1-p+\epsilon)m)+ET(m-1)+O(m^2)
\end{eqnarray}
where the second inequality is due to the fact that $2^{O(m)-\mu m^2}$ is bounded by a constant for all
positive integers $m$.

We then show that $ET(m) \leq 2^{c\log_{2}^{2}{m}}$, where $c$ is a positive constant. We show this by
induction. First, for a sufficiently large positive integer $m_0$ whose value will be specified later, we let
$c_0=\max_{1 \leq t \leq m_0}{\{\frac{\log_{2}{ET(t)}}{\log_{2}^{2}{t}}\}}$ and choose $c=\max{\{c_0,
\frac{2}{\log_{2}{\frac{1}{1-p+\epsilon}}}, 1\}}$. It is not difficult to see that $ET(l) \leq
2^{c\log_{2}^{2}{l}}$ if $1
\leq l \leq m_0$. We then assume this holds for all positive
integers less than $m$. From the above recursion relation on $ET(m)$, we can obtain
\begin{eqnarray}
ET(m) & \leq & 2^{c\log_{2}^2{((1-p+\epsilon)m)}}+2^{c\log_{2}^{2}{(m-1)}}+Bm^2
\\
      & \leq &
      sm^{-l}2^{c\log_{2}^{2}{m}}+2^{c\log_{2}^{2}{m}}+(2^{c\log_{2}^{2}{(m-1)}}-2^{c\log_{2}^{2}{m}})+Bm^2
      \\
      & \leq &
      sm^{-l}2^{c\log_{2}^{2}{m}}+2^{c\log_{2}^{2}{m}}-\frac{\log_{2}{m}}{24m}2^{c\log_{2}^{2}{m}}+Bm^2\\
      & \leq & 2^{c\log_{2}^{2}{m}}
\end{eqnarray}
where $B$ is a positive constant independent of $c, p, \epsilon$ and $s$, $q$, $l$ are some positive constants
that depend on $c, p,
\epsilon$ only. The first inequality is obtained from the assumption
for induction. The second one is due to the fact that
$\log_{2}^2{((1-p+\epsilon)m)}=\log_{2}^2{(1-p+\epsilon)}+2\log_{2}{(1-p+\epsilon)}\log_{2}{m}+\log_{2}^{2}{m}$ and
we can let $l=2c\log_{2}{\frac{1}{1-p+\epsilon}}$ , $s=2^{c\log_{2}^2{(1-p+\epsilon)}}$.

To establish the third inequality, we have
\begin{eqnarray}
\log_{2}^{2}{(m-1)}-\log_{2}^{2}{m} & = & (\log_{2}{m}+\log_{2}{(1-\frac{1}{m})})^2-\log_{2}^{2}{m} \\
                                  & \leq & (\log_{2}{m}-\frac{1}{6m})^2-\log_{2}^{2}{m} \\
                                  & \leq & -\frac{\log_{2}{m}}{6m} \\
                                  & \leq & -\frac{\log_{2}{m}}{6cm}
\end{eqnarray}
when $m \geq 16$, we can obtain
\begin{eqnarray}
2^{c\log_{2}^{2}{(m-1)}}-2^{c\log_{2}^{2}{m}} & = & 2^{c\log_{2}^{2}{m}}(2^{c(\log_{2}^{2}{(m-1)}-\log_{2}^{2}{m})}-1) \\
                                             & \leq & 2^{c\log_{2}^{2}{m}}(2^{-\frac{\log_{2}{m}}{6m}}-1) \\
                                             & \leq & -\frac{\log_{2}{m}}{24m}2^{c\log_{2}^{2}{m}}
\end{eqnarray}
the third inequality thus follows.

From the fact that $c \geq
\frac{2}{\log_{2}{\frac{1}{1-p+\epsilon}}}$, we have $l \geq 4$.
We let
\begin{eqnarray}
c' & = & \frac{2}{\log_{2}{\frac{1}{1-p+\epsilon}}} \\
s' & = & 2^{c'\log_{2}^2{((1-p+\epsilon)m)}} \\
l' & = & 2c'\log_{2}{\frac{1}{1-p+\epsilon}}
\end{eqnarray}
we now consider the function
$F(m)=(s'{m}^{-l'}-\frac{\log_{2}{m}}{24m})2^{c'\log_{2}^{2}{m}}+B{m}^{2}$. Since
$s'$, $l'$, $c'$, and $B$ are independent of $m$ and $l' \geq 4$,
there exists a positive integer $m_1(p,\epsilon)$  such that
$F(m) \leq 0$ when
$m \geq m_{1}(p, \epsilon)$. $m_0$ can be determined as follows
\begin{equation}
m_0=\max\{m_{1}(p, \epsilon), \frac{1}{\sqrt{1-p+\epsilon}}, 16\}
\end{equation}

It is not difficult to see that when $c \geq c'$ and $m \geq m_0$,
we have $s'm^{-l'}-\frac{\log_{2}{m}}{24m} \leq 0$. In addition, we can further verify that
\begin{equation}
sm^{-l} = 2^{c\log_{2}{(1-p+\epsilon)}\log_{2}{(m^2(1-p+\epsilon))}}
\end{equation}
since $c \geq c'$, $m \geq \frac{1}{\sqrt{1-p+\epsilon}}$, and $ \log_{2}{(1-p+\epsilon)} \leq 0$, we can
immediately obtain
\begin{eqnarray}
sm^{-l} &=& 2^{c\log_{2}{(1-p+\epsilon)}\log_{2}{(m^2(1-p+\epsilon))}} \\
        & \leq & 2^{c'\log_{2}{(1-p+\epsilon)}\log_{2}{(m^2(1-p+\epsilon))}} \\
        & = & s'm^{-l'}
\end{eqnarray}
the following thus holds
\begin{eqnarray}
(sm^{-l}-\frac{\log_{2}{m}}{24m})2^{c\log_{2}^{2}{m}}+B{m}^{2} & \leq & (s'm^{-l'}-\frac{\log_{2}{m}}{24m})2^{c\log_{2}^{2}{m}}+B{m}^{2} \\
                                                              & \leq & (s'm^{-l'}-\frac{\log_{2}{m}}{24m})2^{c'\log_{2}^{2}{m}}+Bm^2 \\
                                                              & = & F(m) \\
                                                              & \leq & 0 \\
\end{eqnarray}
the fourth inequality thus follows. From the principle of induction, the theorem has been proved.
\end{proof}
\end{theorem}

\section{Parameterized Algorithm for Independent Set Problem}
The parameterized independent set problem is to decide whether a given graph $G=(V,E)$ contains an independent
set of size $k$ or not. The problem is known to be W[1]-hard \cite{downey,downey1,downey2} and cannot be
solved in time $n^{o(k)}$ in general graphs unless W[2]=FPT \cite{chen,chen2}. We show that if the underlying
graph $G$ is a random graph, the problem can be solved in expected time $2^{O(k^2)}+O(n^3)$, where $n$ is the
number of vertices in the graph. We need the following lemma to analyze the time complexity of the algorithm.
\begin{lemma}
\rm
\label{lm2}
Given a random graph $G=(V, p)$ where $n=|V|$ and a sufficiently small constant $\epsilon$ such that
$p+\epsilon<1$, there exists vertex $u \in V$ such that $deg_{G}(u) \leq (p+\epsilon)n$ with a probability of
at least $1-2^{-\mu n^{2}}$, where $\mu$ is a positive constant that only depends on $\epsilon$ and $p$,

\begin{proof}
The proof is similar to the proof of Lemma \ref{lm1}. If such a vertex does not exist, the degree of every
vertex in $G$ is at least $(p+\epsilon)n$. The graph thus contains at least $\frac{(p+\epsilon)n^2}{2}$ edges.
The expected number of edges in $G$ is $\frac{pn(n-1)}{2}$. We use $n(E)$ to denote the number of the edges in
$G$. From Chernoff bound, we can bound the probability for $G$ to contain at least $\frac{(p+\epsilon)n^2}{2}$
edges.
\begin{eqnarray}
Pr(n(E) & \geq  & \frac{(p+\epsilon)n^2}{2}) \\
        &  < & \exp{(-\frac{\epsilon^{2} n^{2}}{64p})} \\
        & = & 2^{-\frac{\epsilon^{2}n^{2}}{64p\ln{2}}}
\end{eqnarray}
the lemma immediately follows by letting $\mu=\frac{\epsilon^{2}}{64p\ln{2}}$.
\end{proof}
\end{lemma}

The proof of Lemma \ref{lm2} relies on the fact that $p$ is a constant independent of $n$, the Lemma
does not hold if the value of $p$ depends on $n$.

\begin{theorem}
\rm
\label{th2}
Given a random graph $G=(V, p)$, there exists an algorithm that can decide whether $G$ contains an independent
set of size $k$ in expected time $2^{O(k^{2})}+O(n^3)$.

\begin{proof}
We start the proof by comparing the values of $k$ and $L(n)=\frac{1}{3}\log_{\frac{1}{1-p-\epsilon}}{n}$, if
$k>L(n)$, we can enumerate all possible vertex subsets of size $k$ in $G$ and check whether one of them is an
independent set of size $k$ or not. The enumeration and checking needs at most $O(k^{2}n^{k})$ time. However,
since $k> L(n)$, we can obtain $n < (\frac{1}{1-p-\epsilon})^{3k}$, the computation time needed to determine
whether $G$ contains an independent set of size $k$ or not is thus at most
$O(k^{2}(\frac{1}{1-p-\epsilon})^{3k^2})=2^{O(k^2)}$ in this case.

We then consider the case where $k \leq L(n)$. We use the following procedure to generate an independent set
$I$. We start with the vertex $u$ with the minimum degree in $G$, we include $u$ in $I$ and remove $u$ and all
its neighbors in $G$ from $G$. We denote the resulting graph by $G_1$. The procedure can be repeatedly
executed until there are at most $n^{\frac{2}{3}}$ vertices left in the graph. We use $G_0=G, G_1, G_2, G_3,
\cdots, G_l$ to denote the intermediate graphs generated during this iterative procedure. It is not difficult
to see that vertices in $I$ form an independent set in $G$.

We show that the above procedure can generate an independent set $I$ of size at least $L(n)$ with high
probability. We use $G_1, G_2, G_3, \cdots, G_l$ to denote the resulting graph in each iterative step and
$n(G_i)$ to denote the number of vertices in graph $G_i$. From Lemma \ref{lm2}, the following holds with a
probability of at least 1-$2^{-\mu n^{2}(G_i)}$ for each $i$ between $0$ and $l$.
\begin{equation}
n(G_{i+1}) \geq (1-p-\epsilon)n(G_{i})
\end{equation}
Since $n(G_i) > n^{\frac{2}{3}}$, the probability for this inequality to hold for all $i$'s between $0$ and
$l$ is at least $1-n2^{-\mu n^{\frac{4}{3}}}$. If this inequality holds for all $i$'s between $0$ and $l$. We
can immediately obtain
\begin{eqnarray}
l & \geq &
\log_{\frac{1}{1-p-\epsilon}}{(\frac{n}{n^{\frac{2}{3}}})} \\
  & = & \frac{1}{3}\log_{\frac{1}{1-p-\epsilon}}{n} \\
  & = & L(n)
\end{eqnarray}

$I$ thus contains at least $L(n)$ vertices. With a probability of at least $1-n2^{-\mu n^{\frac{4}{3}}}$, the
above iterative procedure generates an independent set of size $L(n)$. Since $k<L(n)$, the algorithm returns
``yes'' if $I$ indeed contains $L(n)$ independent vertices, otherwise, the algorithm simply enumerates all
vertex subsets in $G$ and checks whether one of them is an independent set of size at least $k$. Since the
procedure for generating $I$ needs $O(n^3)$ time, the expected computation time needed for this is at most
\begin{equation}
O(n^{3})(1- n2^{-\mu n^{\frac{4}{3}}})+2^{O(n)}n2^{-\mu n^{\frac{4}{3}}} = O(n^{3})
\end{equation}
where the equality is due to the fact that the second term is bounded by a constant when $n$ is
sufficiently large. The
algorithm thus needs an expected time $2^{O(k^2)}+O(n^3)$, the theorem has been proved.
\end{proof}
\end{theorem}

\section{The Largest Common Subgraph Problem}

Given two graphs $G$, $H$, a {\it common subgraph} of $G$ and $H$ is a third graph $K$ such that both $G$ and
$H$ contain an induced subgraph that is isomorphic to $K$. The largest common subgraph problem is to compute
a common subgraph that contains the largest number of vertices. The problem has important applications in
computational biology. For example, it is often desirable to identify common subgraphs in the protein
interaction networks of two homologous organisms since proteins in these common subgraphs often together play
important roles for certain biological functions \cite{kuchaiev}.

Unfortunately, the problem is NP hard when both of the underlying graphs are general graphs \cite{garey}. The
asymptotically best known algorithm for this problem needs time $O^{*}((m+1)^{n})$ \cite{abu,suters} and
little progress has been made to improve the asymptotical time complexity of this problem. We show that, given
two random graphs $G$ and $H$ in $n$ and $m$ vertices, where $n \geq m$, the largest common subgraph problem
in $G$ and $H$ can be computed in expected time $2^{O(n^{\frac{1}{2}}\log_{2}^{\frac{5}{3}}{n})}$.

\begin{lemma}
\label{lm3}
\rm
The largest common subgraph problem can be solved in computation time $O(m^{2}2^{m}n^{m}) \leq
2^{hm\log_{2}{n}}$, where $h$ is some positive constant that does not depend on $n$ or $m$.

\begin{proof}
We can solve the largest common subgraph problem with the following simple algorithm. For each positive
integer $l$ not greater than $m$, we enumerate all vertex subsets that contain $l$ vertices in $G$. For each
such vertex subset $S_1$, we enumerate all vertex subsets of size $l$ in graph $H$ and for each such vertex
subset $S_2$, we enumerate all possible one to one mappings between vertices in $S_1$ and those in $S_2$. We
then check whether there exists a one to one mapping that can establish the isomorphism between the subgraph
induced by $S_1$ in $G$ and the subgraph induced by $S_2$ in $H$. The algorithm can find all common subgraphs
and return one that is of the largest size.

The number of vertex subsets of size $l$ in $G$ is ${n
\choose l}$ and the number of vertex subsets of size $l$ in $H$ is
${m \choose l}$. The number of one to one mappings between $S_1$ and $S_2$ is $l!$ and the computation time
needed to check whether the two subgraphs induced by $S_1$ and $S_2$ are isomorphic under a particular mapping
is at most $O(l^2)$. The total computation time needed to find and return the largest common subgraph is thus
at most
\begin{eqnarray}
\sum_{l=1}^{m}{C{n \choose l}{m \choose l}l!l^{2}} & \leq & \sum_{l=1}^{m}{Cn^{m}{m \choose l}m^{2}} \\
                                                   & \leq & C2^{m}n^{m}m^{2} \\
                                                   &  \leq & 2^{hm\log_{2}{n}}
\end{eqnarray}
where $C$ and $h$ are some positive constants independent of $n$ and $m$. The first inequality is due to the
fact that ${n \choose l}l!
\leq n^l$ and $l \leq m$; the second inequality is due to the fact
that $\sum_{l=1}^{m}{m \choose l}=2^{m}-1$. The lemma thus has been proved.
\end{proof}
\end{lemma}

\begin{lemma}
\label{lm4}
\rm
Given two random graphs $G=(V, p)$ and $H=(U, q)$, where $p$ and $q$ are positive constants between 0 and 1, $G$
contains $n$ vertices and $H$ contains $m$ vertices ($n \geq m$), the probability that $G$ and $H$ contain a
common subgraph of size $n^{\frac{1}{2}}\log_{2}^{\frac{2}{3}}{n}$ is at most $2^{-\mu
n\log_{2}^{\frac{4}{3}}{n}}$, where $\mu$ is a positive constant that only depends on $p$ and $q$.

\begin{proof}
We let $k=n^{\frac{1}{2}}\log_{2}^{\frac{2}{3}}{n}$ and consider two given subsets of size $k$ in graph $G$
and $H$ respectively. We use $S_1=\{g_1, g_2, \cdots, g_{k}\}$ and $S_2=\{h_1, h_2, \cdots, h_{k}\}$ to denote
them and $G_1$, $H_1$ to denote the subgraphs induced by them in $G$ and $H$ respectively. We assume that
$G_1$ is isomorphic to $H_1$ under a given one to one mapping $M$, where vertex $g_i$ in $S_1$ is mapped to
$h_i$ in $S_2$ for $1 \leq i
\leq k$.

We then estimate the probability for $M$ to be such a mapping. If $G_1$ is isomorphic to $H_1$ under $M$, for
any integer pair $(i, j)$, where $1 \leq i < j \leq k$, either both $(g_i, g_j)$ and $(h_i, h_j)$ are edges or
neither of them are edges. The probability for the former case is $pq$ and the probability for the latter case is
$(1-p)(1-q)$. Since there are in total $\frac{k(k-1)}{2}$ such pairs, the probability for $G_1$ and $H_1$ to
be isomorphic under $M$ is thus $(pq+(1-p)(1-q))^{\frac{k(k-1)}{2}}$.

We use $P(k)$ to denote the probability for $G$ and $H$ to contain a common subgraph of size $k$. Since the
number of vertex subsets of size $k$ in $G$ is ${n \choose k}$ and the number of vertex subsets of size $k$ in
$H$ is ${m \choose k}$, we can obtain an upper bound for $P(k)$ using the union bound.
\begin{eqnarray}
P(k) & \leq & {n \choose k}{m \choose k}\sum_{M}{s^{\frac{k(k-1)}{2}}} \\
     & \leq & {n \choose k}{m \choose k}k!s^{\frac{k(k-1)}{2}} \\
     & \leq & n^{k}m^{k}s^{\frac{k(k-1)}{2}} \\
     & \leq & n^{2k}s^{\frac{k(k-1)}{2}} \\
     &  \leq & 2^{2n^{\frac{1}{2}}\log_{2}^{\frac{5}{3}}{n}}s^{\frac{k^2}{4}} \\
     &  = &
     2^{2n^{\frac{1}{2}}\log_{2}^{\frac{5}{3}}{n}-\frac{n\log_{2}^{\frac{4}{3}}{n}\log_{2}{\frac{1}{s}}}{4}}
     \\
     &  \leq & 2^{-\mu n\log_{2}^{\frac{4}{3}}{n}}
\end{eqnarray}
where $s=pq+(1-p)(1-q)$ and $\mu$ is some positive constant that depends on $p$ and $q$ only. The first
inequality is due to the union bound. The second inequality follows from the fact that there are in total $k!$
one to one mappings between vertices in $S_1$ and $S_2$. The third inequality is due to the fact that ${n
\choose k}
\leq n^{k}$ and ${m \choose k} k! \leq m^{k}$. The fifth inequality
follows from the fact that $k=n^{\frac{1}{2}}\log_{2}^{\frac{2}{3}}{n}$ and $\frac{k(k-1)}{2}
> \frac{k^2}{4}$ when $n$ is sufficiently large. The last inequality
is due to the fact that $s<1$ and $2n^{\frac{1}{2}}\log_{2}^{\frac{5}{3}}{n} \leq
\frac{n\log_{2}^{\frac{4}{3}}{n}\log_{2}{\frac{1}{s}}}{8}$ for sufficiently large $n$.
\end{proof}
\end{lemma}

The proof of the Lemma \ref{lm4} relies on the fact that $p$ and $q$ are
both constants independent of $n$ and $m$, the Lemma does not hold if the values of $p$ and $q$ depend on
$n$ or $m$.

\begin{theorem}
\rm
\label{th4}
Given two random graphs $G=(V, p)$ and $H=(U, q)$, where $p$ and $q$ are positive numbers between 0 and 1, $G$
contains $n$ vertices and $H$ contains $m$ vertices ($m \leq n$), a largest common graph of $G$ and $H$ can be
computed in expected time $2^{O(n^{\frac{1}{2}}\log_{2}^{\frac{5}{3}}{n})}$.

\begin{proof}
We only need to show that such an algorithm exists when $m > n^{\frac{1}{2}}\log_{2}^{\frac{2}{3}}{n}$. Since
if $m \leq n^{\frac{1}{2}}\log_{2}^{\frac{2}{3}}{n}$, the algorithm in the proof of Lemma \ref{lm3} can be
directly used to find a largest common subgraph of $G$ and $H$ in time
$2^{O(n^{\frac{1}{2}}\log_{2}^{\frac{5}{3}}{n})}$.

Let $k=n^{\frac{1}{2}}\log_{2}^{\frac{2}{3}}{n}$, since $m > k$, we can use the following algorithm to compute
a largest common subgraph in $G$ and $H$.
\begin{enumerate}
\item{Enumerate all vertex subsets of size $k$ in $G$. For
each such vertex subset $S_1$, enumerate all vertex subsets of size $k$ in $H$;}
\item{for each such subset $S_2$ in $H$, we enumerate all possible one to one mappings
between $S_1$ and $S_2$;}
\item{for each such mapping $M$, determine whether the subgraph induced by $S_1$ in $G$
is isomorphic to the subgraph induced by $S_2$ in $H$ under $M$ or not;}
\item{if there exists a mapping that can make the subgraph induced by $S_1$ in $G$
isomorphic to the subgraph induced by $S_2$ in $H$, call the algorithm in Lemma \ref{lm3} to compute a
largest common subgraph of $G$ and $H$ and return it;}
\item{otherwise, for each integer $i$ between $1$ and $k$, use the same approach
as described in steps 1, 2, 3 to determine whether $G$ and $H$ contains a common subgraph of size $i$ or
not;}
\item{Based on the result of the exhaustive search performed in step $5$,
return a common subgraph of the largest size.}
\end{enumerate}

We then show that the algorithm can compute the largest common subgraph of $G$ and $H$ in expected
$2^{O(n^{\frac{1}{2}}\log_{2}^{\frac{5}{3}}{n})}$ time. In particular, the computation time needed by the
exhaustive search performed in steps 1, 2, and 3 is at most
\begin{eqnarray}
C{n \choose k}{m \choose k}k!k^{2} & \leq & Cn^{k}m^{k} \\
                                   & \leq & Cn^{2k} \\
                                   &  \leq & C2^{2k\log_{2}{n}} \\
                                   & = & 2^{O(n^{\frac{1}{2}}\log_{2}^{\frac{5}{3}}{n})}
\end{eqnarray}
where $C$ is some positive constant. The first inequality is due to the fact that ${m \choose k}k! \leq m^{k}$
and ${n \choose k}k^{2} \leq n^{k}$ for sufficiently large $n$. From Lemma \ref{lm3}, step 4 of the algorithm,
if executed, needs $2^{hm\log_{2}{n}}$ computation time, where $h$ is some positive constant independent of
$n$ and $m$. The computation time needed by step 5 is at most
\begin{eqnarray}
D\sum_{i=1}^{k-1}{{n \choose i}{m \choose i}i!i^{2}} & \leq & D\sum_{i=1}^{k-1}{n^{i}m^{i}i^{2}} \\
                                                       & \leq & Dkn^{2k}k^{2} \\
                                                       & = & D2^{3\log_{2}{k}+2k\log_{2}{n}} \\
                                                       &  = & 2^{O(n^{\frac{1}{2}}\log_{2}^{\frac{5}{3}}{n})}
\end{eqnarray}
where $D$ is some positive constant. The first inequality is due to the fact that ${m \choose i} i! \leq
m^{i}$ and ${n \choose i} \leq n^{i}$. The second inequality follows from the fact that $1 \leq i < k$.

Only one of steps 4 and 5 is executed by the algorithm. From Lemma
\ref{lm4}, the probability for step 4 to be executed is at most
$2^{-\mu n\log_{2}^{\frac{4}{3}}{n}}$, where $\mu$ is some constant that depends on $p$ and $q$ only. The
expected computation time to execute steps 4 and 5 is thus at most
\begin{equation}
2^{O(n^{\frac{1}{2}}\log_{2}^{\frac{5}{3}}{n})}+2^{hm\log_{2}{n}}2^{-\mu n\log_{2}^{\frac{4}{3}}{n}}.
\end{equation}
Since $m \leq n$, the second term is bounded by a constant for sufficiently large $n$. We thus can conclude
that the expected computation time for steps 4 and 5 is $2^{O(n^{\frac{1}{2}}\log_{2}^{\frac{5}{3}}{n})}$.
Since steps 1, 2, 3 also need $2^{O(n^{\frac{1}{2}}\log_{2}^{\frac{5}{3}}{n})}$ computation time, the theorem
has been proved.
\end{proof}
\end{theorem}

\section{Approximate Algorithm}

As discussed in the introduction,
the maximum independent set problem cannot be approximated within a ratio of $n^{1-\epsilon}$
in polynomial time unless P=NP, where $\epsilon$ is any positive constant.
In \cite{boppana}, it is shown that the maximum independent set in a graph can be approximated within a ratio
of $O(\frac{n}{\log_{2}^{2}{n}})$. In \cite{fiege}, the approximation ratio is improved to
$O(\frac{n\log_{2}^{2}{\log_{2}{n}}}{\log_{2}^{3}{n}})$. The result so far remains the best known
approximation ratio achieved for this problem in general graphs. In \cite{grimmett,homer,oghlan}, a polynomial
time algorithm that can approximate the maximum independent set in a random graph within a constant ratio with
high probability is developed and analyzed. However, the approximation ratio of the algorithm is not
guaranteed to be constant for all graphs. We show that, the maximum independent set in a random graph can be
approximated within a ratio of $\frac{2n}{2^{\sqrt{\log_{2}{n}}}}$ in expected polynomial time, which is a
significant improvement compared with the best known approximate ratio for this problem in general graphs.

\begin{theorem}
\rm
Given a random graph $G=(V,p)$ in $n$ vertices where $p$ is a positive constant between $0$ and $1$, the
maximum independent set in $G$ can be approximated within a ratio of $\frac{2n}{2^{\sqrt{\log_{2}{n}}}}$ in
expected polynomial time.

\begin{proof}
We use the following simple algorithm to compute an independent set in $G$. We let $k=\lfloor
2^{\sqrt{\log_{2}{n}}} \rfloor$ and partition the vertices in $G$ into $l$ disjoint vertex subsets such that
$l-1$ of them contains $k$ vertices and the remaining one contains at most $k$ vertices. We use $G_{1}, G_{2},
\cdots, G_{l}$ to denote the subgraph induced by vertices in these vertex subsets. It is not difficult to see
that $l \leq \lfloor \frac{n}{k}
\rfloor +1$.

We then use the algorithm we have developed in Theorem \ref{th1} to compute a maximum independent set in each
of $G_1, G_2, \cdots, G_l$ and return the one that contains the largest number of vertices.

We first show that the algorithm returns an independent set in expected polynomial time. $G_1, G_2, \cdots,
G_l$ are disjoint and the expected time needed to compute a maximum independent set in each of them is at most
$2^{c\log_{2}^{2}{k}}$, where $c$ is some positive constant that only depends on $p$. Since $k \leq
2^{\sqrt{\log_{2}{n}}}$, the expected computation time needed to compute the maximum independent set in one
subgraph is at most $2^{c\log_{2}{n}}=n^{c}$. The algorithm thus returns an independent set in expected time
$n^{c+1}$.

We then show that the algorithm can achieve an approximate ratio of $\frac{2n}{2^{\sqrt{\log_{2}{n}}}}$. We
use $APX(G)$ to denote the size of the independent set returned by the algorithm and $OPT(G)$ to denote the
size of a maximum independent set in $G$. we assume that $I$ is a maximum independent set in $G$. Since we
have partitioned the graph $G$ into $l$ disjoint subgraphs $G_0, G_1,
\cdots, G_l$, at least one of the $l$ subgraphs contains at least
$\frac{OPT(G)}{l}$ vertices from $I$. These vertices form an independent set in the subgraph. Since the
algorithm computes a maximum independent set in each subgraph and returns the one with the largest size, we
immediately obtain
\begin{equation}
APX(G) \geq \frac{OPT(G)}{l}
\end{equation}
this suggests that
\begin{eqnarray}
\frac{OPT(G)}{APX(G)} & \leq & l   \\
                      & \leq & \lfloor \frac{n}{k} \rfloor+1 \\
                      & \leq & \frac{n}{k}+1 \\
                      & \leq & \frac{n}{2^{\sqrt{\log_{2}{n}}}-1}+1 \\
                      & \leq & \frac{2n}{2^{\sqrt{\log_{2}{n}}}}.
\end{eqnarray}
The second inequality is due to the fact that $l \leq
\lfloor \frac{n}{k} \rfloor+1$. The fourth inequality is due to the fact
that $k \geq 2^{\sqrt{\log_{2}{n}}}-1$. The last inequality holds for sufficiently large $n$. The theorem thus
has been proved.
\end{proof}
\end{theorem}

\section{Conclusions}

In this paper, we study the independent set problem in random graphs. We show that a maximum independent set
in a random graph can be computed in expected subexponential time. We also show that the parameterized
independent set problem is fixed parameter tractable with high probability for random graphs. Using
techniques based on enumeration, we show that the largest common subgraph in two random graphs can be computed in expected
subexponential time. Our work also suggests that the maximum independent set in a random graph can be
approximated within a ratio of $\frac{2n}{2^{\sqrt{\log_{2}{n}}}}$ in expected polynomial time, which
significantly improves on the best known approximate ratio for this problem in general graphs.

It remains unknown whether the maximum independent set in a random graph can be computed in expected
polynomial time or not. One possible direction of future work is to study whether there exists such an
algorithm. Another related open question is that if such an algorithm does not exist, whether it can be
approximated within an improved ratio in expected polynomial time. Further investigations are needed to solve
these problems.


\begin{thebibliography}{25}
\bibliographystyle{plain}
\bibitem{abu}
F. N. Abu-Khzam, N. F. Samatova, M. A. Rizk, and M. A. Langston, ``The Maximum Common Subgraph Problem: Faster
Solutions via Vertex Cover'', {\it Proceedings of 2007 IEEE/ACS International Conference on Computer Systems
and Applications (AICCSA 2007)}, pp. 367-373, 2007.
\bibitem{balas}
E. Balas and C. S. Yu,``Finding a Maximum Clique in An Arbitrary Graph'', {\it SIAM Journal on Computing},
15(4):1054-1068, 1986.
\bibitem{battiti}
R. Battiti and M. Protasi, "Reactive Local Search for the Maximum Clique Problem", {\it Algorithmica} 29(4):
610-637, 2001.
\bibitem{boppana}
R. Boppana and M. Halld\'{o}rson,``Approximating Maximum Independent Sets by Excluding Subgraphs'', {\it BIT
Computer Science and Numerical Mathematics},32(2):180-196, 1994.
\bibitem{brinda}
K. V. Brinda, S. Vishveshwara and S. Vishveshwara, ``Random Network Behaviour of Protein Structures'', {\it
Molecular, BioSystems}, 6:391-398, 2010.
\bibitem{carraghan}
R. Carraghan and P. M. Pardalos, ``An Exact Algorithm for the Maximum Clique Problem'', {\it Operations
Research Letters}, 9(6): 375-382, 1990.
\bibitem{chen}
J. Chen, X. Huang, I. A. Kanj, and G. Xia, ``Linear FPT Reductions and Computational Lower Bounds'', {\it
Proceedings of the Thirty-Sixth ACM Symposium on Theory of Computing (STOC 2004)}, pp.212-221, 2004.
\bibitem{chen2}
J. Chen, X. Huang, I. A. Kanj, and G. Xia, ``Strong Computational Lower Bounds via Parameterized Complexity'',
{\it Journal of Computer and System Sciences}, 72(8):1346-1367, 2006.
\bibitem{dinur}
I. Dinur and S. Safra, ``The Importance of Being Biased'', {\it Proceeding of the Thirty-Fourth ACM Symposium
on Theory of Computing (STOC 2002)}, pp. 33-42, 2002.
\bibitem{downey}
R. G. Downey and M. R. Fellows, {\it Parameterized Complexity}, Springer-Verlag, 1998.
\bibitem{downey1}
R. G. Downey and M. R. Fellows, ``Fixed Parameter Tractability and Completeness i: Basic Theory'', {\it SIAM
Journal of Computing}, 24:873-921, 1995.
\bibitem{downey2}
R. G. Downey and M. R. Fellows, ``Fixed Parameter Tractability and Completeness ii: Completeness for W[1]'',
{\it Theoretical Computer Science A}, 141:109-131, 1995.
\bibitem{erdos}
P. Erd\H{o}s and A. R\'{e}nyi, ``On Random Graphs'', {\it Publicationes Mathematicae}, 6: 290-297, 1959.
\bibitem{fahle}
T. Fahle, ``Simple and Fast: Improving a Branch-And-Bound Algorithm for Maximum Clique'', {\it Proceedings of
the Tenth European Symposium on Algorithms} pp. 47-86, 2002.
\bibitem{fiege}
U. Fiege, ``Approximating Maximum Clique by Removing Subgraphs'', {\it SIAM Journal on Discrete Mathematics},
18(2):219-225, 2004.
\bibitem{fiege2}
U. Fiege and E. Ofek,``Finding A Maximum Independent Set in A Sparse Random Graph'', {\it SIAM Journal on
Discrete Mathematics}, 22(2):693-718, 2008.
\bibitem{fomin}
F. V. Fomin, F. Grandoni, and D. Kratsch, ``Measure and Conquer: A simple $O(2^{0.288n})$ Independent Set
Problem'', {\it Proceedings of the Seventeenth ACM-SIAM Symposium on Discrete Algorithms (SODA 2006)}, pp.
18-25, 2006.
\bibitem{garey}
M. R. Garey and D. S. Johnson, {\it Computers and Intractability}, W. H. Freeman and Co., San Francisco,
California, 1979. A guide to the theory of NP-completeness, A Series of Books in the Mathematical Sciences.
\bibitem{grimmett}
G. R. Grimmett and C. J. H. Mcdiarmid, ``On Colouring Random Graphs'', {\it Mathematical Proceedings of the
Cambridge Philosophical Society}, 77(2):313-324, 1975.
\bibitem{grosso}
A. Grosso, M. Locatelli, F. D. Croce, ``Combining Swaps and Node Weights in An Adaptive Greedy Approach for
the Maximum Clique Problem'', {\it Journal of Heuristics} 10(2):135-152, 2004.
\bibitem{hastad}
J. H\r{a}stad, ``Clique Is Hard to Approximate Within $n^{1-\epsilon}$'', {\it Proceedings of the 37th Annual
Symposium on Foundations of Computer Science (STOC 1996)}, 627-636, 1996.
\bibitem{homer}
S. Homer and M. Peinado, ``On the performance of Polynomial-time CLIQUE Approximation Algorithms on Very Large
Graphs'', {\it In Cliques, Coloring, and Satisfiability: second DIMACS Implementation Challenge}, pp. 103-124,
1993.
\bibitem{jian}
T. Jian, ``An $O(2^{0.308n})$ Algorithm for Solving Maximum Independent Set Problem'', {\it IEEE Transactions
on Computers}, 35(9):847-851, 1986.
 \bibitem{johnson}
D. S. Johnson, ``Approximate Algorithms for Combinatorial Problems'', {\it Journal of Computer and System
Sciences}, 9, 256-278, 1974.
\bibitem{karp}
R. M. Karp, ``The Probability Analysis of Some Combinatorial Search Problems'', {\it Algorithms and
Complexity: New Directions and Recent Results}, 1-19, Academic Press, New York, 1976.
\bibitem{katayama}
K. Katayama, A. Hamamoto, and H. Narihisa, ``An Effective Local Search for the Maximum Clique Problem'', {\it
Information Processing Letters} 95(5):503-511, 2005.
\bibitem{konc}
J. Konc and D. Jane\v{z}i\v{c}, ``An Improved Branch and Bound Algorithm for the Maximum Clique Problem'',
{\it MATCH Communications in Mathematical and in Computer Chemistry} 58(3): 569-590, 2007.
\bibitem{kuchaiev}
O. Kuchaiev, T. Milenkovi\'{c}, V. Memi\v{s}evi\'{c}, W. Hayes, and N. Pr\v{z}ulj, ``Topological Network
Alignment Uncovers Biological Function and Phylogeny'', {\it Journal of Royal Society Interface},
7(50):1341-1354, 2010.
\bibitem{oghlan}
A. Coja-Oghlan and C. Efthymiou, ``On Independent Sets in Random Graphs'', {\it Proceedings of the Twenty
Second Annual ACM-SIAM Symposium on Discrete Algorithms (SODA 2011)}, pp. 136-144, 2011.
\bibitem{ostergard}
P. R. J. \"{O}sterg\r{a}rd, ``A Fast Algorithm for the Maximum Clique Problem'', {\it Discrete Applied
Mathematics} 120 (1–3):197-207,2002.
\bibitem{pardalos}
P. M. Pardalos and G. P. Rogers, ``A Branch and Bound Algorithm for the Maximum Clique Problem'', {\it
Computers and Operations Research} 19 (5): 363-375, 1992.
\bibitem{regin}
J. C. R\'{e}gin, ``Using Constraint Programming to Solve the Maximum Clique Problem'', {\it Proceedings of the
Ninth International Conference on Principles and Practice of Constraint Programming}, pp. 634-648, 2003.
\bibitem{robson}
J. M. Robson, ``Algorithms for Maximum Independent Sets'', {\it Journal of Algorithms}, 7(3):425-440, 1986.
\bibitem{robson1}
J. M. Robson, ``Finding A Maximum Independent Set in Time $O(2^{\frac{n}{4}})$'', {\it Technical Report
1251-01, LaBRI Universit\'{e} de Bordeaux I}, 2001.
\bibitem{suters}
W. H. Suters, F. N. Abu-Khzam, Y. Zhang, C. T. Symons, N. F. Samatova, and M. A. Langston, ``A New Approach
and Faster Exact Methods for the Maximum Common Subgraph Problem'',{\it Proceedings of the Eleventh
International Computing and Combinatorics Conference}, pp. 717-727, 2005.
\bibitem{tarjan}
R. E. Tarjan and A. E. Trojanowski, ``Finding A Maximum Independent Set", {\it Technical Report CS-TR-76-550},
Stanford University, 1976.
\bibitem{tomita}
E. Tomita and T. Seki, ``An Efficient Branch-and-bound Algorithm for Finding a Maximum Clique'', {\it Discrete
Mathematics and Theoretical Computer Science}, pp. 278-289, 2003.
\bibitem{tomita1}
E. Tomita and T. Kameda, ``An Efficient Branch-and-bound Algorithm for Finding A Maximum Clique with
Computational Experiments'', {\it Journal of Global Optimization} 37(1): 95-111, 2007.
\end{thebibliography}
\end{document}